\documentclass[10pt, aps, prd, amssymb, amsmath, amsfonts, showpacs, floatfix, twocolumn, twoside, a4paper, superscriptaddress, longbibliography, nofootinbib]{revtex4-2}
\usepackage[utf8]{inputenc}

\usepackage{amsmath}
\usepackage{amssymb}
\usepackage{bm}
\usepackage{enumitem}
\usepackage[acronym]{glossaries}
\usepackage{graphicx}
\usepackage[colorlinks,unicode]{hyperref}
\usepackage[capitalise]{cleveref}  % must come after hyperref
\usepackage{mathtools}
\usepackage{multirow}
\usepackage{physics}
\usepackage{siunitx}
\usepackage[dvipsnames]{xcolor}
\usepackage{tcolorbox}
\usepackage{todonotes}
\usepackage{microtype} % For nicer alignment of text
\usepackage[normalem]{ulem}
\usepackage{nicefrac}

\DeclareSIUnit\week{week}

\makeatletter
\renewcommand\onecolumngrid{%
\do@columngrid{one}{\@ne}%
\def\set@footnotewidth{\onecolumngrid}%
\def\footnoterule{\kern-6pt\hrule width 1.5in\kern6pt}%
}
\makeatother

\newcommand\myshade{80}
\colorlet{mylinkcolor}{ForestGreen}
\colorlet{mycitecolor}{Red}
\colorlet{myurlcolor}{violet}

\hypersetup{
  linkcolor  = mylinkcolor!\myshade!black,
  citecolor  = mycitecolor!\myshade!black,
  urlcolor   = myurlcolor!\myshade!black,
  colorlinks = true
}

\newcommand{\calE}{\mathcal{E}}

\DeclareSymbolFont{mathtx}{OML}{txmi}{m}{it}
\DeclareMathAlphabet\mathbfcal{OMS}{cmsy}{b}{n}
\DeclareMathSymbol{v}{\mathalpha}{mathtx}{118}

\begin{document}

\title{Mass and spin coevolution of black holes inspiralling through dark matter}

\author{Theophanes K. Karydas}
\email{t.karydas@uva.nl}
\affiliation{\GRAPPA}
\author{Rodrigo Vicente}
\affiliation{\GRAPPA}
\author{Gianfranco Bertone}
\affiliation{\GRAPPA}

\newcommand{\GRAPPA}{Gravitation Astroparticle Physics Amsterdam (GRAPPA),\\ University of Amsterdam, 1098 XH Amsterdam, The Netherlands}

\begin{abstract}
In extreme/intermediate–mass-ratio inspirals (E/IMRIs) embedded in dark-matter (DM) spikes, the secondary black hole can accrete collisionless particles from the surrounding halo. We study how the companion’s spin controls this process, and the ensuing back-reaction on the magnitude and direction of the companion’s spin vector. We find that higher spin suppresses the mass accretion rate but enhances the accretion-induced torques, driving spin-down and secular alignment of the companion’s spin with the orbital plane. Collisionless DM accretion generically imprints a near-universal mass–spin correlation characterized by a spin-evolution parameter \(s \simeq 2.8\), much larger than is the case for typical astrophysical environments, and largely independent of the local DM density and the spike slope. The associated spin-down proceeds on astrophysically relevant timescales, thus observations of rapidly spinning IMRI companions would disfavor the presence of dense DM environments, providing constraints complementary to those arising from dynamical friction.
\end{abstract}

\maketitle

\section{Introduction}
% \vspace*{-0.5em}

The dynamics of black hole (BH) inspirals in vacuum is accurately predicted by general relativity (GR). Next-generation millihertz gravitational wave (GW) detectors, such as LISA \cite{LISA_2024hlh}, 
TianQin \cite{TianQin:2015yph}, and Taiji \cite{Hu:2017mde} will be able to track extreme and intermediate mass-ratio BH binaries over extended durations. These observations will allow us to precisely probe the environments in which binaries evolve~\cite{Barausse:2014tra, Cardoso:2019rou, LISA:2022kgy, CanevaSantoro:2023aol, Bertone:2024rxe, Zwick:2025wkt, Zwick:2025wkt, Cardoso_2022, spieksma2025blackholespectroscopyenvironments, Cole:2022yzw, Roy:2024rhe}, opening new opportunities for discoveries in fundamental physics, from gravity~\cite{Amaro_Seoane_2007, Gair_2013, speri2024probingfundamentalphysicsextreme, Mitra_2024} to particle physics~\cite{PhysRevLett.133.121404, Barack_2019, Bertone_2020, Baumann_2022, Maselli_2022, miller2025gravitationalwaveprobesparticle}.
For instance, dark matter (DM) \emph{spikes} around BHs~\cite{Gondolo_1999, Sadeghian_2013, Ferrer_2017, bertone2024darkmattermoundsrealistic, Ullio_2001, Zhao_2005} can alter the orbital evolution of binaries \cite{Eda_2013, Kavanagh_2020, Coogan_2022, Cole_2023, karydas2024sharpeningdarkmattersignature, kavanagh2024sharpeningdarkmattersignature, Speeney_2022, Becker_2022, mitra2025extrememassratioinspirals, Li_2022, Zhang_2024, Mukherjee_2024, Hannuksela_2020, Becker_2023, zhou2025intermediatemassratioinspiralsgeneraldynamical, Nichols_2023, Edwards_2020, Montalvo_2024, Vicente:2025gsg}, leaving imprints of the microscopic properties of DM in their waveforms. Such signatures can, in principle, be distinguished from the effects of other dense astrophysical environments, such as accretion discs~\cite{Cole:2022yzw}.

The interaction of BH binary systems with (collisionless) particle DM is mediated only by gravity. The leading environmental effects arise from dynamical friction and accretion~\cite{chandra1, chandra2, chandra3, Traykova_2023}. 
Both these effects are sensitive to the internal degrees of freedom of the small companion object (in the BH case, its mass and spin). For instance, it has recently been shown that the spin of a Kerr BH leads to gravitational Magnus and lift forces as it moves through a distribution of collisionless particles~\cite{Dyson:2024qrq} (see also~\cite{Costa_2018, wang2024gravitationalmagnuseffectscalar,mach2025accretionvlasovgaskerr}). 

Here, we study the accretion process onto a spinning BH companion in the case of an extreme/intermediate mass ratio inspiral (E/IMRI)~\cite{Amaro_Seoane_2007, Amaro-Seoane:2012lgq} embedded in a DM spike~\cite{Eda_2013, Kavanagh_2020, Coogan_2022, Cole_2023, karydas2024sharpeningdarkmattersignature, kavanagh2024sharpeningdarkmattersignature, Speeney_2022, Becker_2022, mitra2025extrememassratioinspirals, Li_2022, Zhang_2024, Mukherjee_2024, Hannuksela_2020, Becker_2023, zhou2025intermediatemassratioinspiralsgeneraldynamical, Nichols_2023, Edwards_2020, Montalvo_2024, Vicente:2025gsg}. We show that the accretion of cold and collisionless DM particles within such environments leads to a characteristic coevolution of the companion's mass and spin. While we focus on DM spikes, our framework can be directly applied to any other collisionless system described by a particle distribution function. Similarly, while we focus on (quasi)circular E/IMRIs, our approach can be easily extended to other setups.

The paper is organized as follows. In~\cref{ssec:mass_force} we investigate how DM accretes onto a spinning BH, and calculate how the mass accretion rate~$\dot{m}$ varies with spin for a BH immersed in a collisionless environment. 
In~\cref{ssec:angtum_transfer}, we calculate how the angular momentum carried by the accreted particles alters the spinning state of the Kerr companion. Finally, in~\cref{ssec:spin} we bring together the previous calculations to study how the mass and spin 
of the companion co-evolve over astrophysical timescales, and argue that a measurement of the companion's spin can be a supplementary probe to the existence of DM spikes. We discuss the results in \cref{sec:discussion} and present our conclusions in \cref{sec:conclusions}.
\begin{figure}[t!]
    \centering
    \hspace*{0.03\columnwidth}
    \includegraphics[width=0.78\columnwidth]{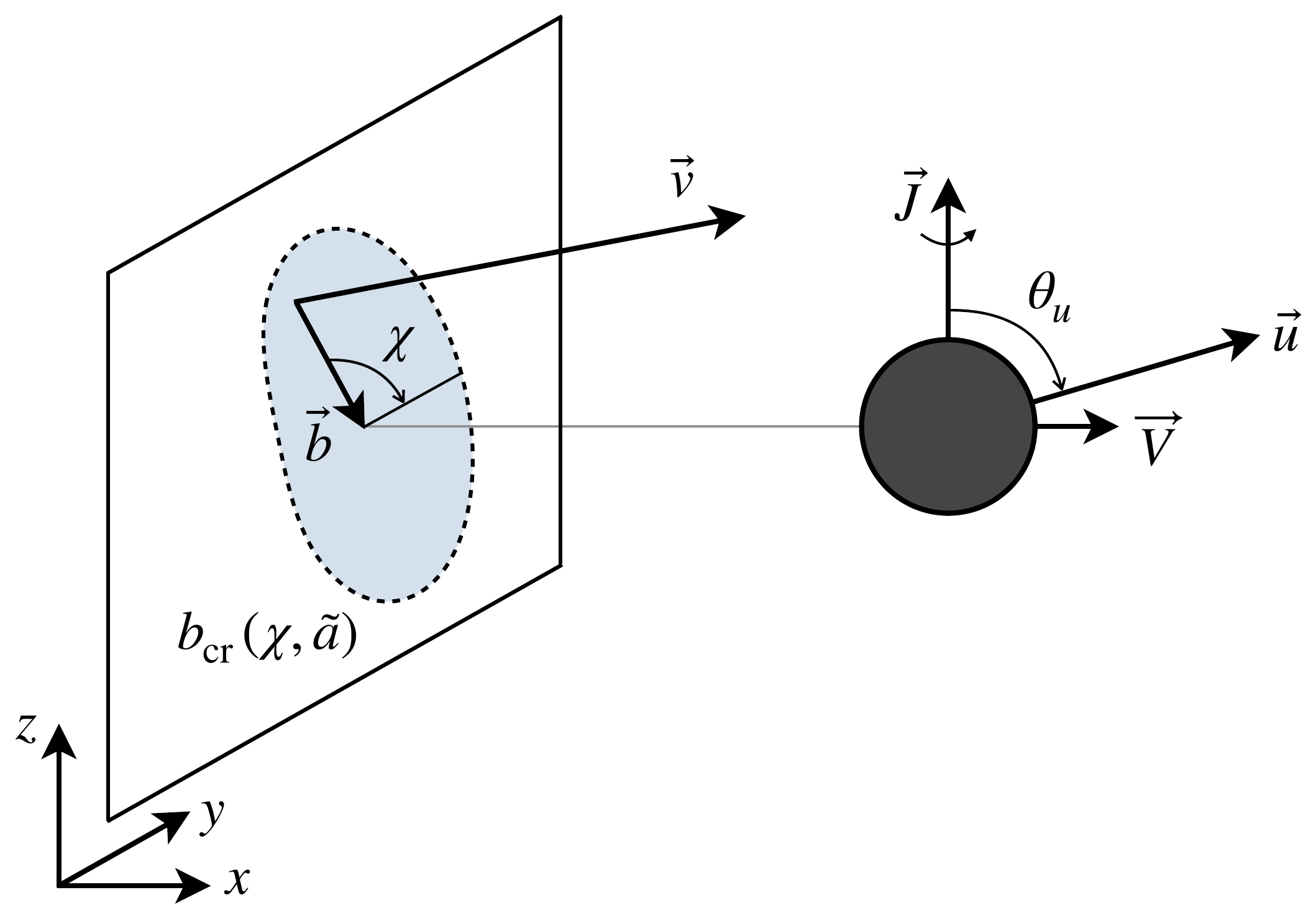}
    \vspace*{-0.5em}
    \caption{\textbf{Geometry of collisionless DM capture by a spinning companion BH.} For a given relative velocity $\bm{V}$, the blue shaded region shows the set of impact parameters with $b\leq b_\mathrm{cr}(\chi,\tilde{a})$, that lead to capture. The capture contour is lopsided, with a larger effective radius for retrograde encounters, here corresponding to $\chi \in (-\pi/2, \pi/2) $. 
    For illustration purposes, we choose $\bm V \parallel \hat{\bm x}$ and place the impact-parameter vector $\bm b$ in the $y$–$z$ plane.
    \label{fig:diagram}}
    \vspace*{-1.5em}
\end{figure}

\vspace*{-1.2em}
\section{Mass accretion within a collisionless DM environment} \label{ssec:mass_force}
\vspace*{-0.2em}

\setlength{\skip\footins}{3em}

Consider a BH with spin $\bm{J}$ aligned along the $z$-axis inspiralling through a DM environment into a much heavier (primary) BH. We aim to determine the flux of captured DM particles and the ensuing transfer of mass and momentum to the smaller BH. As sketched in Fig.~\ref{fig:diagram}, each particle scatters off the BH in its reference frame with impact parameter vector~$\bm{b}$ and relative velocity vector~$\bm{V}$ (with $\theta_V$ and $\phi_V$, respectively, the relative velocity's polar and azimuthal angles). 
The DM environment is described by a distribution function~$f(\bm{V}\!,\bm{r})$, counting the number of particles with position and velocity within a phase space element $\dd^3\bm{V} \dd^3 \bm{r}$ in the BH frame. 
Only particles with impact parameter smaller than the critical $b_{\rm cr}(\chi,V,\theta_V,\tilde{a})$ are captured into the BH, where $\chi$ is the angle of $\bm{b}$ relative to $\bm{\hat{z}}\times\bm{\hat{V}}$ (see Fig.~\ref{fig:diagram}) and
\begin{equation}
    \tilde{a}=cJ/(Gm^2)
\end{equation}
 is the dimensionless Kerr spin parameter. In App.~\ref{app:bcr_calculate}, we show how to find~$b_{\rm cr}$ as a function of the integrals of motion in Kerr geometry, which can then be related to the asymptotic quantities $(\chi, V, \theta_V, \tilde{a})$.

The mass accretion rate is
\begin{equation} \label{eq:accretion_in_kerr}
  \dot{m} = \mu \int \gamma^2 f(\bm{V}\!,\bm{r}) V \sigma_{\rm acc}(\bm{V},\tilde{a}) \, \mathrm{d}^3 \bm{V} \,,
\end{equation}
where~$\mu$ is the rest mass of the DM particles,~$f(\bm{V}\!,\bm{r})$ is evaluated at the position of the accreting small BH, $\gamma^2\equiv 1/(1-V^2/c^2)$, and the accretion cross-section is given by
\begin{equation} \label{eq:csection}
    \sigma_{\rm acc}=\int_0^{2\pi} \frac{b^2_{\rm cr}(\chi,V,\theta_V,\tilde{a})}{2} \mathrm{d}\chi\,.
\end{equation}
In Eq.~\eqref{eq:accretion_in_kerr}, the quadratic power in the Lorentz-factor $\gamma$ originates from the relativistic energy of the DM particles and volume contraction in the companion BH frame. In the zero spin limit, this expression reduces~to~\cite[Eq.~3.8]{karydas2024sharpeningdarkmattersignature}.

The initial particle distribution function is typically known in the primary BH rest frame, and we thus perform the integration over the velocity space in Eq.~\eqref{eq:accretion_in_kerr} in that frame.
We are interested in the long-term evolution of the small BH mass, very early in the inspiral, at large separation distances.
In that regime, both the particle's velocity, $\mathbf{v}$, and the small BH one, $\bm{u}$, are non-relativistic. 
We therefore take $\gamma\approx 1$ and, from the Galilean velocity transformation~$\bm{V} = \mathbf{v} -\bm{u}$, we express
\begin{equation*}
\begin{gathered}
    \sin\theta_V\cos\phi_V = \frac{ \sin\xi\cos\psi\cos\theta_u +\left(\cos\xi-u/v \right)\sin\theta_u}{V/v}\,,
    \\
    \sin\theta_V\sin\phi_V = -\frac{v}{V}\sin\xi\sin\psi \,,
    \\
    \cos\theta_V = \frac{-\sin\xi\cos\psi\sin\theta_u +\left(\cos\xi-u/v \right)\cos\theta_u}{V/v}\,,
\end{gathered}
\end{equation*}
with $V^2 = u^2 +v^2 -2 u v \cos\xi$; here, $\xi$ and $\psi$ are the polar and azimuthal angles of $\mathbf{v}$ with respect to $\bm{u}$. Instead of performing the velocity space integration in the variables $(V,\theta_V,\phi_V)$, it will be more convenient to do it in the variables $(v,\xi,\psi)$. At each instant of time, the velocity of the small BH is described in spherical coordinates by $(u,\theta_u,\phi_u)$. Without loss of generality, we align the spherical coordinates such that $\phi_u=0$, i.e., with the BH velocity in the $x$--$z$ plane (note again that $\bm{\hat{z}}= \bm{\hat{J}}$).

Far from the primary BH, the DM particles are not only moving non-relativistically but also nearly isotropically, as loss-cone effects are negligible in these regions~\cite{Kavanagh_2020, Coogan_2022, Cole_2023, karydas2024sharpeningdarkmattersignature}. Thus, as in~\cite{Kavanagh_2020, Coogan_2022, Cole_2023, karydas2024sharpeningdarkmattersignature}, we can obtain the DM spike distribution function by applying the Eddington inversion procedure (e.g., p.~290 of~\cite{binney}) to a power-law density profile $\rho\propto r^{-\gamma_\mathrm{sp}}$. 
The resulting distribution function is a function of only $v^2$ and $r/R_M$, with $R_M\equiv G M/c^2$ the primary BH's gravitational radius.

\begin{figure}[t]
\centering
\hspace*{-0.08\columnwidth}
    \includegraphics[width=\columnwidth]{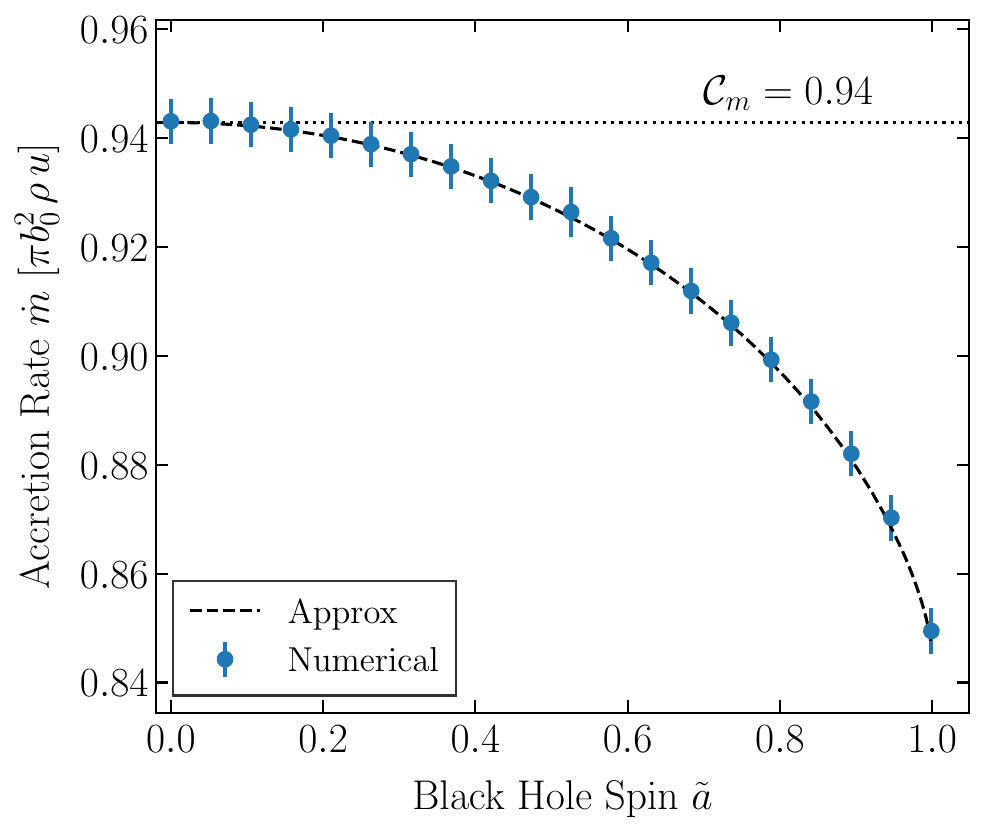}
    \caption{\textbf{Accretion rate onto the companion black hole in an EMRI embedded in a dark matter spike, shown as a function of the companion’s dimensionless spin parameter.
    } The blue points represent the numerical result of the quasi-Monte Carlo integration and associated 95\% confidence intervals. The black dashed~line depicts \cref{eq:approx_results}, while the dotted line shows the same expression but for a BH with $\tilde{a}=0$ (for comparison).
    The results are for an EMRI/IMRI at separation $r=10^3 R_M$ in a DM spike with $\gamma_\mathrm{sp} = 7/3$~\cite{Gondolo_1999}, and an angle between BH's velocity and spin-axis 
    $\theta_u = \pi/3$.
\label{fig:accretion_mass_force}}
\end{figure}

Under these assumptions, we integrate \cref{eq:accretion_in_kerr} numerically using a quasi-Monte Carlo (qMC) scheme~\cite{quasi} on a low-discrepancy sequence~\cite{SOBOL196786}, with a linear matrix scrambling and digital random shift~\cite{OWEN1998466}. We also represent the integration error arising from the finite number of qMC samples at the~$95\%$ confidence intervals obtained via the bootstrapping method~\cite{efron1979bootstrap}. The results are shown in \cref{fig:accretion_mass_force}, which depicts the dependence of the accretion rate on the small BH's spin magnitude.

Our results show that the BH spin suppresses the accretion rate by $7$--$10\%$ for a maximally spinning BH compared to the non-rotating case, with the suppression being the largest when the spin-axis is perpendicular to that of the BH's velocity.\\

\paragraph*{\textbf{An approximate relationship.}}

Alternatively, we can obtain an approximate expression for the accretion rate by recasting it in terms of quantities that depend on the velocity of the BH, $u$, supplemented by a correction factor to account for the particle distribution. This form is simpler to evaluate and can then be used to explore the accretion rate across the parameter space. For that, we first note that~\cref{eq:accretion_in_kerr} can be rewritten as
\begin{equation*}
    \dot{m} = \rho \,u\, \sigma_{\rm acc}(u,\theta_u,\tilde{a}) \, \mathcal{C}_m(r,u,\theta_u,\tilde{a}) \,,
\end{equation*}
where $\rho(r)$ is the DM density at the companion's location, and $\mathcal{C}_m$ is a velocity-distribution correction factor~\cite{karydas2024sharpeningdarkmattersignature}
\begin{equation} \label{eq:cm}
    \mathcal{C}_m\equiv \int \frac{\sigma_{\rm acc}(V,\theta_V,\tilde{a}) }{\sigma_{\rm acc}(u,\theta_u,\tilde{a})} \frac{V}{u} \frac{\mu f(\mathbf{v}, r)}{\rho(r)} \dd^3\mathbf{v} \,.
\end{equation}
Our numerical exploration indicates that $\mathcal{C}_m(r,u,\theta_u,\tilde{a}) \approx \mathcal{C}_m(r,u,\tilde{a}=0)$, implying that the accretion rate is well-approximated by (cf.~\cref{fig:accretion_mass_force})
\begin{equation}\label{eq:approx_results}
    \dot{m} \approx \rho(r) \, u\, \sigma_{\rm acc}(u,\theta_u,\tilde{a}) \, \mathcal{C}_m(r,u)\,.
\end{equation}
The dependence of the accretion rate on $\theta_u$ and $\tilde{a}$ originates primarily from $\sigma_{\rm acc}(u,\theta_u,\tilde{a})$.

For a circular orbit of radius $r=10^3 R_M$ in a DM spike environment of slope $\gamma_\mathrm{sp} = 7/3$~\cite{Gondolo_1999}, we find $\mathcal{C}_m\approx 0.94$. For circular orbits, this value is nearly independent of the radius for $r\gtrsim 50 R_M$, and changes only slightly with $\gamma_{\rm sp}\in \left( 1.5, 3\right)$; the latter dependence is well described by $\mathcal{C}_m\approx0.94+0.05(\gamma_{\rm sp}-7/3)$ up to an error $<1.5\%$.

\section{Spin evolution within a collisionless DM environment} \label{ssec:angtum_transfer}

Matter accretion not only increases the mass of the small BH but also modifies the magnitude and orientation of its spin. 
Conservation of angular momentum gives the per-encounter transfer
\(\Delta\bm{J}=\mu\,\bm{b}\!\cross\!\bm{V}\).
Integrating over impact parameters and velocities of particles drawn from
\(f(\bm{V},\bm{r})\) (as in the previous section) yields
\begin{equation} \label{eq:angtum_change}
    \dot{\bm{J}} = \mu \int \left[\int b\,  (\bm{b} \cross \bm{V}) \dd b \, \dd \chi \right] \gamma^2  f(\bm{V}\!,\bm{r}) V \mathrm{d}^3 \bm{V} \,.
\end{equation}

We (again) focus on early-stage E/IMRIs at large separations, where the velocity distribution in the primary BH frame is nearly isotropic and concentrated at non-relativistic speeds ($\gamma\approx 1$).
Expressing the angular momentum in \cref{eq:angtum_change} in terms of the impact plane angle $\chi$, \linebreak $\bm{b} \cross \bm{V} = -bV (\bm{\hat{\theta}_V} \cos \chi + \bm{\hat{\phi}_V} \sin \chi )$,%
\footnote{
An incoming particle approaches with an impact parameter vector $\bm{b} = b\, (\bm{\hat{\theta}_V}\sin \chi  -\bm{\hat{\phi}_V} \cos \chi)$, where
$\bm{\hat{\theta}_V} = \bm{\hat{x}} \cos\theta_V\cos\phi_V +\bm{\hat{y}}\cos\theta_V \sin\phi_V  -\bm{\hat{z}} \sin\theta_V$ and $\bm{\hat{\phi}_V} = -\bm{\hat{x}} \sin\phi_V +\bm{\hat{y}}\cos\phi_V$.
} and integrating over the impact parameters, we find
\begin{equation} 
  \dot{\bm{J}} = -\frac{\mu}{3} \int \left[\int b_{\rm cr}^3(\chi, \bm{V}) \cos\chi \mathrm{d}\chi\right] \bm{\hat{\theta}_V} V^2 f\, \mathrm{d}^3 \bm{V} \,,
\end{equation}
where we have trivially evaluated the contribution parallel to $\bm{\hat{\phi}_V}$ which vanishes, because $b_{\rm cr}$ is a function of $\chi$ only through $\cos\chi$. We use the trigonometric relations of the previous section to perform the integration over the velocity space in the primary BH's frame. As expected, note that $\dot{\bm{J}}$ vanishes entirely when $\tilde{a}=0$, as the critical impact parameter $b_{\rm cr}$ becomes independent of $\chi$. 

The rate from \cref{eq:spin_djdt} can be decomposed into two parts: a contribution that alters the magnitude of the angular momentum $\dot{J}_{z} \equiv \dot{\bm{J}} \cdot \bm{\hat{z}}$; and a torque which is perpendicular to the spinning axis and alters the spin's direction. The BH's velocity along $\bm{\hat{y}}$ vanishes (by construction) and, so, from the symmetry of the problem, it is evident that after integrating over the azimuthal angle $\psi$ the torque's projection on the $y$-axis vanishes, $\dot{J}_y\equiv\dot{\bm{J}}\cdot \bm{\hat{y}}=0$. The resulting rate of change is
\begin{equation} \label{eq:spin_djdt}
 \bm{\dot J} = -\frac{\mu}{3} \int\int b_{\rm cr}^3 \cos \chi \mathrm{d}\chi
 { \begin{pmatrix}
   \cos\theta_V \cos\phi_V \\ 0 \\ -\sin\theta_V
 \end{pmatrix}}
  V^2 f \, \mathrm{d}^3 \mathbf{V}. 
\end{equation}

\begin{figure}[t!]
\centering
\hspace*{-0.1\columnwidth}
    \includegraphics[width=\columnwidth]{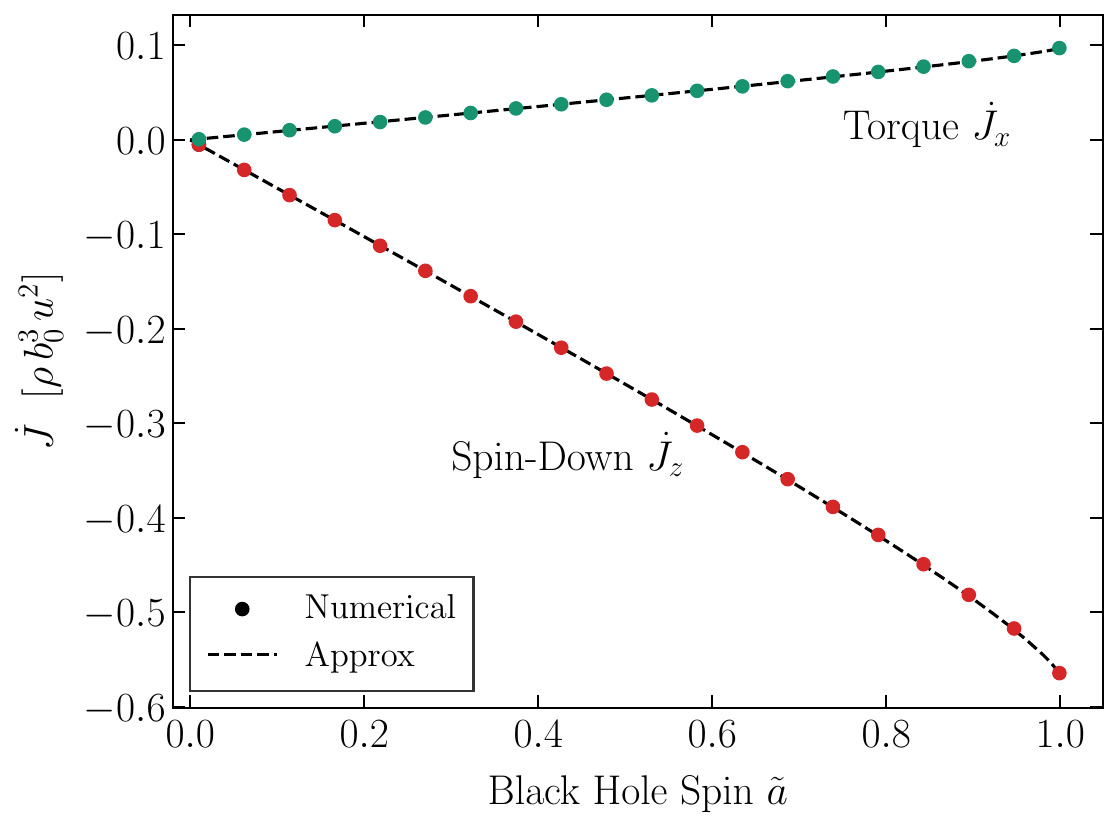}
    \caption{\textbf{Rate of change of the companion’s spin as a function of its spin parameter for an EMRI in a dark matter spike.
} The colored dots denote the numerical results for the same configuration as in~\cref{fig:accretion_mass_force}. The black dashed lines show the simpler approximate expression from \cref{eq:dJdt_approx}.\label{fig:torque_alpha}}
\end{figure}

Our results are shown in~\cref{fig:torque_alpha}, which depicts the two non-vanishing components of $\dot{\bm{J}}$ and their dependence on the BH's (instantaneous) spin for the same configuration as in~\cref{fig:accretion_mass_force}. The component along $\bm{\hat{z}}$ describes the rate of change of the spin's magnitude, while the component along $\bm{\hat{x}}$ describes instead the change in the spin-axis direction. 
We find that collisionless DM environments cause BHs to spin down, and that this process is more efficient for higher BH spins, due to the critical impact parameter $b_{\rm cr}$ being larger for retrograde than for prograde encounters.  

Our numerical analysis also reveals that~$\dot{J}_x \propto \sin{2 \theta_u}$, indicating three critical spin-axis configurations where $\dot{J}_x=0$: one where the BH's spin is orthogonal to its velocity ($\theta_u = \pi/2$), and the others where the two are (anti)aligned ($\theta_u = 0,\, \pi$).\\

\paragraph*{\textbf{An approximate relationship.}}

We conclude this section by adopting the same strategy as before to derive simple approximate expressions for the components of $\dot{\bm{J}}$. 
We start by recasting the torque from DM accretion as
\begin{align*} 
    \begin{pmatrix}
    \dot{J}_x \\ \dot{J}_z
  \end{pmatrix} &=  \rho \, u^2 \sigma_J(u, \theta_u, \tilde{a}) \, \mathbfcal{C}_J(r, u, \theta_u, \tilde{a})\,,
\end{align*}
with the geometric \emph{volume} factor
\begin{equation} \label{eq:sigmaJ}
    \sigma_J(V, \theta_V, \tilde{a}) \equiv -\frac{1}{3} \int b^3_{\rm cr}(\chi, V, \theta_V, \tilde{a}) \cos\chi \, \mathrm{d} \chi\,,
\end{equation}
which can be seen as an effective cross-section for angular momentum transfer (c.f., \cref{eq:csection}) and vanishes identically for $\tilde a = 0$ (from spherical symmetry), and with the velocity-distribution factor
\begin{equation} \label{eq:cj0}
    \mathbfcal{C}_J = \!\!\int \!\frac{
    \sigma_J(\bm{V}, \tilde{a})}{\sigma_J(\bm{u}, \tilde{a})} \frac{V^2}{u^2}\!
 \begin{pmatrix}
   \cos\theta_V \cos\phi_V \\ -\sin\theta_V
 \end{pmatrix}
 \frac{\mu f(\mathbf{v,r})}{\rho(r)} \, \mathrm{d}^3 \mathbf{v}.
\end{equation}

Our numerical exploration indicates that $\mathbfcal{C}_J$ is well approximated by the simple $\tilde{a}$-independent form
\begin{equation}
    \mathbfcal{C}_J(r,u,\theta_u) \approx - \frac{1}{l_1(u) \sin\theta_u}
    \begin{pmatrix}
        -\mathcal{C}_x \sin\left( 2 \theta_u\right) \\ \mathcal{C}_z + 2\,\mathcal{C}_x \sin^2 \theta_u
    \end{pmatrix},
\end{equation}
where the function $l_1(u)$ is defined in \cref{app:expansions}; the denominator $F_1 \equiv l_1 \sin\theta_u$ arises as the leading order term in the spin expansion of \cref{eq:sigmaJ}, as shown in \cref{app:expansions}. This results in the approximation
\begin{align} \label{eq:dJdt_approx}  
    \begin{pmatrix}
    \dot{J}_x \\ \dot{J}_z
  \end{pmatrix} &\approx  \rho \, u^2 \sigma_J(u, \theta_u, \tilde{a}) \, \mathbfcal{C}_J(r, u, \theta_u)\,,
\end{align}
where the dependence of the torque on $\tilde{a}$ originates primarily from $\sigma_J(u, \theta_u, \tilde{a})$. In the following sections, we will further expand $\sigma_J$ in $\tilde{a}$, as shown in \cref{app:expansions}, to derive (semi)analytic results.

The factors $\mathcal{C}_x$ and $\mathcal{C}_z$ are obtained by fitting to the numerical evaluation of \cref{eq:cj0}. For a circular orbit of radius $r= 10^3 R_M$ in a DM spike of slope $\gamma_\mathrm{sp} = 7/3$~\cite{Gondolo_1999}, we find $\mathcal{C}_x \approx 0.101(3)$ and $\mathcal{C}_z \approx 0.359(7)$. As was the case for $\mathcal{C}_m$ in the previous section, for circular orbits these coefficients are nearly independent of $r\gtrsim 50 R_M$, and vary only slightly with $\gamma_{\rm sp}\in \left( 1.5, 3\right)$; the latter dependence is captured by $\mathcal{C}_x\approx 0.10+0.05 (\gamma_{\rm sp}-7/3)$ and \linebreak $\mathcal{C}_z\approx 0.36-0.04 (\gamma_{\rm sp}-7/3)$ up to an error $<1.5\%$.

\section{Mass and spin coevolution on astrophysical timescales} \label{ssec:spin}

As we have shown above, the spin vector of BHs evolves (both in direction and magnitude) as their mass grows by accreting from a collisionless DM environment. In this section, we start by analyzing the evolution of the spin-axis tilt over astrophysical timescales, considering both isolated BHs and companions in E/IMRIs. We then do the same for the spin magnitude; here, there is no qualitative distinction between isolated BHs and E/IMRI. We then assess the
joint effect on the \emph{effective spin} of the companion---the spin component orthogonal to the orbital plane---which is the better constrained parameter through GW observations.

To give a scale for the time associated with the evolution of the small BH's mass and spin for typical DM spike environments, in the following sections, we will consider the following fiducial parameters for an IMRI within a DM spike environment (e.g., \cite{karydas2024sharpeningdarkmattersignature}):
\begin{equation}\label{eq:fiducial}
\begin{gathered}
    M=10^4 \, {\rm M}_\odot\,,\qquad m=10 \, {\rm M}_\odot\, ,\\ 
    \gamma_\mathrm{sp} = 7/3\,, \qquad \rho(10^{-6}\,\mathrm{pc}) = 10^{16}\,\mathrm{M}_\odot\, \mathrm{pc}^{-3}\,.
\end{gathered}
\end{equation}
These parameters serve to provide a sense of the typical scales; however, our framework and qualitative conclusions are general.
The choice of our binary parameters is also motivated by: (i) the survival of the spike to hierarchical mergers being more robust around intermediate mass BHs~\cite{Zhao_2005, Bertone:2005xz}, and (ii) the effective spin being better constrained through the GW observations of less extreme mass-ratios~\cite{Huerta:2011zi, Piovano_2021}. Extrapolation of our results to ``golden EMRIs'' for mHz detectors will be discussed in \cref{sec:discussion}. 

\subsection{Evolution of the spin-axis} \label{ssec:precession}
\paragraph*{\textbf{Isolated BHs in linear motion.}}
For these systems, it is instructive to examine the evolution of the spin-axis through the angle between the BH's spin and the velocity, $\theta_u$, whose rate of change is given by
\begin{equation}\label{eq:dottheta}
    \dot{\theta}_u \equiv -\frac{1}{\sin \theta_u}\frac{\mathrm{d}}{\mathrm{d}t}\left(\bm{\hat  J} \cdot \bm{\hat u}\right)
    = - \frac{c\, \dot{J}_x}{Gm^2 \tilde a}  - \frac{ \dot{\bm{\hat u}} \cdot \bm{\hat z}}{\sin\theta_u}\,.
\end{equation}
The second term vanishes for linear motion, $\dot{\bm{\hat u}} = 0$.

The three critical spin orientations where the torque $\dot{J}_x$ vanishes become equilibrium configurations. In particular, asymmetric accretion makes the (anti)parallel configurations (with $\theta_u = 0,\, \pi$) attractor solutions over the characteristic timescale $\tau_\theta$,\footnote{The solution $\cot\theta_u\left(t\right) = \cot[\theta_{u}(0)] e^{2t/\tau_\theta}$ shows that (formally) equilibrium is only achieved as $t \to \infty$; in practice, $\theta_u \approx 0$ or $\theta_u \approx\pi$ is attained in $t\sim\mathcal{O}(\tau_\theta)$.}
\begin{equation}
    \dot{\theta}_u \approx -\frac{\sin\left(2\theta_u\right)}{\tau_\theta}\,, \!\quad \text{where} \quad \!\tau_\theta^{-1} \equiv \frac{c}{Gm^2} \rho(r)\, u^2 b^3_\mathrm{cr} \, \mathcal{C}_x \,. \nonumber
\end{equation}
The last expression is valid up to $\mathcal{O}(\tilde{a}^3)$ and follows from substituting \cref{eq:dJdt_approx} into \cref{eq:dottheta}, and expanding $\sigma_J$ using~\cref{eq:fitb3}. Conversely, the other equilibrium configuration, $\theta_u = \pi/2$, is unstable. From \cref{eq:approx_results}, and by expanding $\sigma_\mathrm{acc}$ using~\cref{eq:fitb2}, we find that the accretion process leads to a correlation between the companion's spin orientation and its mass,
\begin{equation} \label{eq:lin-corelation}
    \frac{\mathrm{d}\theta_u}{\mathrm{d}\ln m} \approx - \frac{4}{\pi} \frac{\mathcal{C}_x}{\mathcal{C}_m} \sin\left(2\theta_u\right)\,,
\end{equation}
which is independent of the local environment density. \\

\paragraph*{\textbf{Orbiting BH companions.}} 
For a BH companion orbiting a massive central BH within a DM spike, the spin-axis dynamics is qualitatively different. Such dynamics depends on the ratio between the orbital period \linebreak $T_\mathrm{orb} = 2\pi r/u$ (for a circular orbit), and the spin-axis reorientation timescale~$\tau_\theta$,
\begin{equation}
    \frac{T_\mathrm{orb}}{\tau_\theta} = q\frac{128\pi \mathcal{C}_x G}{c^2} r^2\rho(r) \,.
\end{equation}
Specifically,
\begin{itemize}[label=-]
    \item  $T_\mathrm{orb}/\tau_\theta \gg 1$ : the system reaches equilibrium near instantaneously, and the axis (anti)aligns with the orbital velocity, as in the isolated case.
    \item $T_\mathrm{orb}/\tau_\theta \lesssim 1$: the spin-axis reorientation lags behind the orbital velocity change, evolving in a complex way.
    \item $T_\mathrm{orb}/\tau_\theta \ll 1$: the torque from collisionless DM accretion secularly aligns the spin-axis parallel to the orbital plane.
\end{itemize}

The last regime characterizes typical DM environments, even for densities as large as those in DM spikes. 
For the fiducial values in \cref{eq:fiducial} and and orbital separation $r = 100 R_M$, we obtain $T_\mathrm{orb}/\tau_\theta \approx 5 \times 10^{-11}$. Comparable values are found across the realistic parameter space.

To see that for $T_\mathrm{orb}/\tau_\theta \ll 1$ the spin-axis is secularly aligned parallel to the orbital plane, consider the following. 
First, we note that the angle between the spin-axis and velocity, $\theta_u$, is related to the true anomaly, $\nu$, and the angle between the spin-axis and orbital angular momentum, $\theta_L$, through $\cos \theta_u = - \sin \theta_L \sin \nu$; in particular, for circular orbits, $\nu = \omega t$, where $\omega$ is the (orbital) angular velocity. Second, due to the hierarchy of timescales $T_\mathrm{orb}/\tau_\theta \ll 1$, it will be convenient to take orbit averages $\langle \cdots \rangle \equiv T_\mathrm{orb}^{-1} \int_{T_\mathrm{orb}} (\cdots) \, \dd t$ to extract the secular evolution of the spin-axis orientation. From $\langle \dot{J}_x\rangle =0$, \cref{{eq:dottheta}} implies that $\langle \dot{\theta}_u \rangle=0$, i.e., the spin-axis will not align with the velocity vector (since the latter revolves too rapidly).

Following the same approach that led us to \cref{eq:dottheta}, we obtain that the spin tilt, $\theta_L$, changes at a rate
\begin{equation} \label{eq:dthetaL_full}
    \dot{\theta}_L \equiv -\frac{1}{\sin\theta_L} \frac{\mathrm{d}}{\mathrm{d}t} \left( \bm{\hat{J}} \cdot \bm{\hat{L}_\mathrm{orb}} \right) = - \frac{\sin\theta_u}{\cos\theta_L \, \sin\nu} \frac{\dot{J}_x}{J}\,.
\end{equation}
Interestingly, the orbit average of the right-hand side of the last equation does not vanish. In fact, using \cref{eq:dJdt_approx} with $\sigma_J$ expanded up to $\mathcal{O}(\tilde{a}^3)$ (c.f., \cref{eq:fitb3}), we obtain
\begin{equation} \label{eq:dthetaL}
    \langle\dot{\theta}_L\rangle = \frac{1}{\tau_\theta} \tan\theta_L \left(2 - \sin^2\theta_L \right) + \mathcal{O}(\tilde{a}^3) \,.
\end{equation}
This shows that the spin-axis is secularly aligned parallel to the orbital plane, i.e., $\theta_L \to \pi/2$. This attractor can be understood from the fact that the effective cross-section $\sigma_J$ is larger for on-axis encounters.
The accretion dynamics leads to a correlation between the companion's spin tilt and its mass; from \cref{eq:approx_results}, and by expanding $\sigma_\mathrm{acc}$ using~\cref{eq:fitb2}, we find
\begin{equation} \label{eq:dthetaLdlnm_approx}
    \frac{\mathrm{d}\theta_L}{\mathrm{d}\ln m}  =  \frac{4\mathcal{C}_x}{\pi \mathcal{C}_m} \tan\theta_L \left( 2 -\sin^2 \theta_L\right) +\mathcal{O}(\tilde a^2)\,,
\end{equation}
which, surprisingly, is independent of the local environment density and the orbital parameters.

\begin{figure}[t!]
\centering
\hspace*{-0.05\columnwidth}
    \includegraphics[width=0.9\columnwidth]{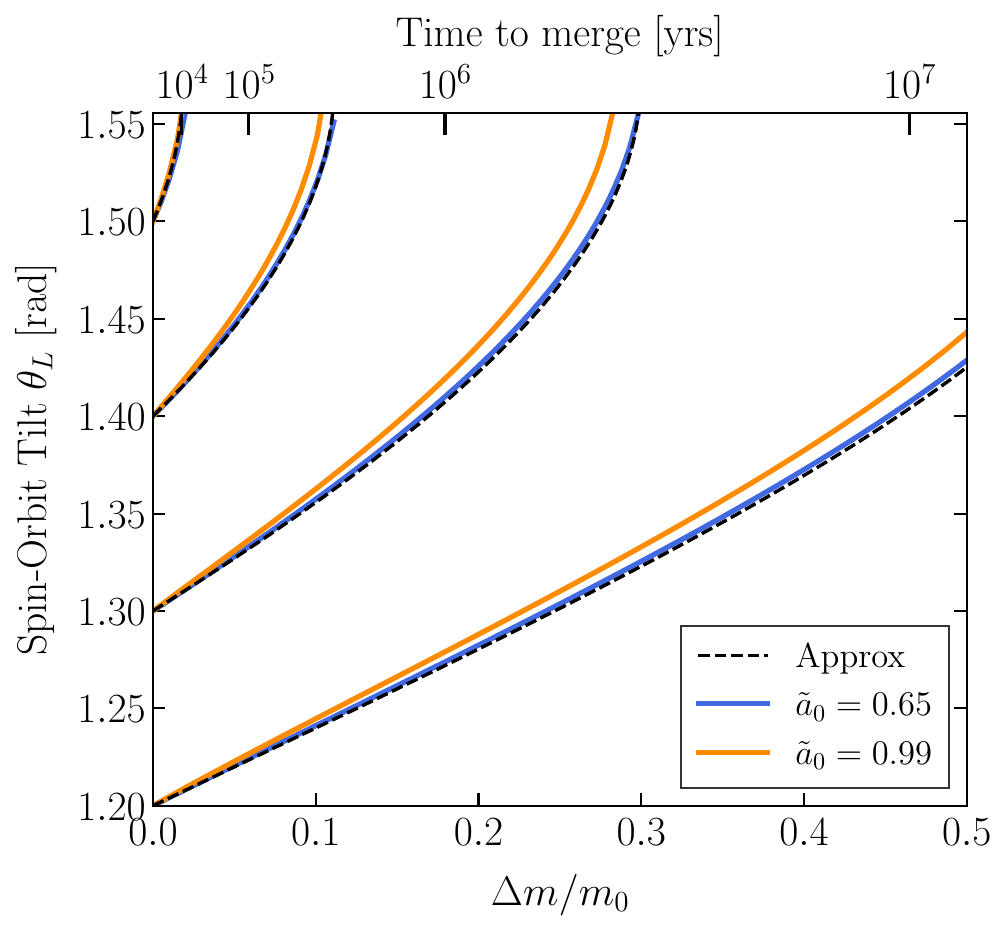}
    \caption{\textbf{Evolution of the spin–orbit tilt $\bm{\theta_L}$ as a function of the fractional mass growth $\bm{\Delta m/m_0}$.} Solid lines show the full numerical integration of the coupled evolution of spin tilt and magnitude [from \cref{eq:dthetaL_full,eq:dadt}, together with \cref{eq:approx_results}]. Dashed lines correspond to the (leading order in $\tilde{a}$) analytic solution in \cref{eq:approx_theta}. At the top, we indicate the time to merge, assuming a quasi-circular inspiral driven by GW emission and accretion drag for the fiducial parameters \eqref{eq:fiducial}.\label{fig:thetaL}}
\end{figure}

In \cref{fig:thetaL}, we present the coevolution of the spin-orbit tilt---the angle between the spin-axis and the orbital angular momentum---of a companion BH in quasi-circular orbit with its mass. We show the results obtained by numerically integrating $\mathrm{d}\theta/\mathrm{d}m = \langle\dot{\theta}\rangle/\langle\dot{m}\rangle$ using \cref{eq:approx_results,eq:dthetaL_full}, and considering the coupled evolution of the spin magnitude $\tilde{a}$ from \cref{eq:dadt} (discussed in detail in the next section). We also compare the results to the (leading order in $\tilde{a}$) solution of \cref{eq:dthetaLdlnm_approx}, which has the analytic form
\begin{equation} \label{eq:approx_theta}
    \theta_L(m) \approx \arcsin\Bigl[\tfrac{1}{2}+\bigl(\csc^{2}\theta_L(m_0)-\tfrac{1}{2}\bigr)\bigl(\tfrac{m}{m_0}\bigr)^{-\frac{16\mathcal{C}_x}{\pi \mathcal{C}_m}}\Bigr]^{-\frac{1}{2}}\,.
\end{equation}
This approximation neglects terms $\mathcal{O}(\tilde a^2)$; yet, as shown in~\cref{fig:thetaL}, it is an excellent approximation for $\tilde{a} \lesssim 0.7$, with relative error smaller than $\sim 3\%$ even for spins as large as $\tilde{a}\approx0.99$. We emphasize that the tilt evolution is effectively decoupled from the spin magnitude; this decoupling is explicit at $\mathcal{O}(\tilde{a}^2)$ (as seen from \cref{eq:approx_theta}).

To provide a timescale for the alignment process, we present our results accompanied by the associated time to merge, assuming a quasi-circular inspiral driven by GW emission and accretion drag (implemented as in~\cite{karydas2024sharpeningdarkmattersignature}), with the fiducial parameters in \cref{eq:fiducial}.

\subsection{Evolution of the spin magnitude} \label{ssec:magnitude}
We now turn to the corresponding change in spin magnitude. The rate of change of the spin parameter is~\cite{abramowic,thorne,Hughes_2003}
\begin{align} \label{eq:dadt}
    \dot{\tilde{a}} &= - 2\tilde{a}\, \frac{\dot{m}}{m} +\frac{c\, \dot{J}_z}{Gm^2}\,,
\end{align}
where the first term describes a BH spin-down (at fixed angular momentum) from its mass growth alone, while the second accounts for the additional angular momentum brought in by the accreted DM particles.

It will be convenient to introduce the (dimensionless) spin-evolution parameter
\begin{equation} \label{eq:s}
    s \equiv -\frac{d \ln \tilde{a}}{d \ln m} = 2 - \frac{\dot{J}_z \, c}{G m \dot{m} \tilde {a}}\,,
\end{equation}
which is often used in the literature (e.g., \cite{2004ApJ...602..312G,King_2008}). In the absence of a net angular momentum transfer (e.g., spherical accretion), the spin-evolution parameter is $s=2$. Importantly, the last term is sensitive to the accretion dynamics---and so to the BH astrophysical environment (e.g., \cite{10.1046/j.1365-8711.1999.02482.x,ricarte2023recipesjetfeedbackspin,McKinney_2004,2005ApJ...620...59S})---through the ratio $\dot{J}_z/\dot{m}$.
A spin-evolution parameter $s < 2$ corresponds to a positive flux of angular momentum flowing into the BH, while $s > 2$ corresponds to a negative flux instead.

\begin{figure}[ht!]
\centering
\hspace*{-0.05\columnwidth}
    \includegraphics[width=1.03\columnwidth]{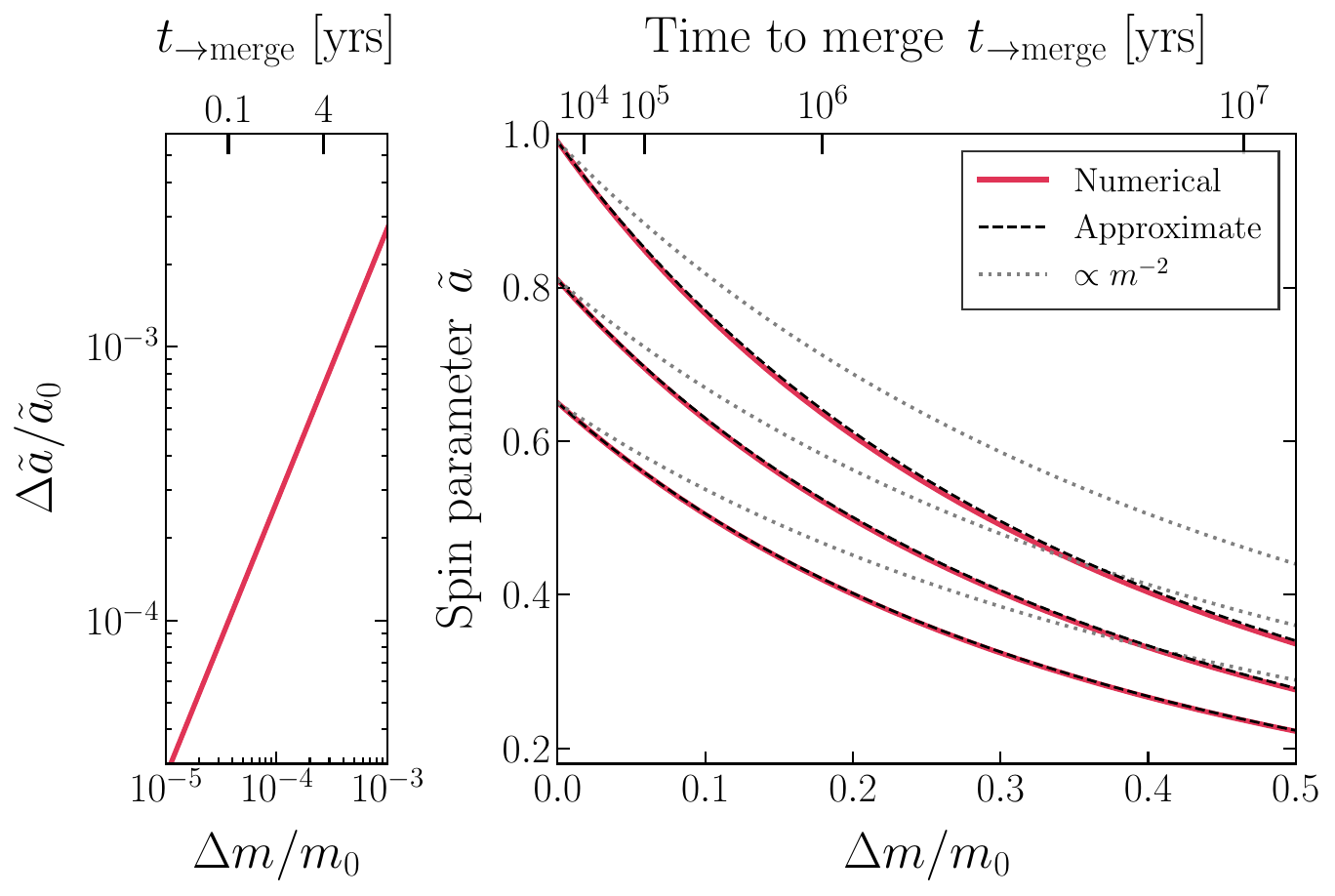}
    \caption{\textbf{Evolution of the spin parameter $\bm{\tilde{a}}$ as a function of the fractional mass growth $\bm{\Delta m/m_0}$.} On the right: solid lines represent $\tilde{a}(\Delta m)$ from numerically integrating \cref{eq:s}, dashed lines are the (leading order in $\tilde{a}$) solutions of \cref{eq:s_bound}, and dotted lines denote the $s=2$ solution. On the left: the fractional change in spin during the last few years before the merger. All results are for the fiducial parameters \cref{eq:fiducial} and a spin-orbit tilt $\theta_L = \pi/2$.\label{fig:alpha_mdot}}
\end{figure}

For a DM spike environment, the parameter $s$ can be computed from \cref{eq:approx_results,eq:dJdt_approx}. At leading order in $\tilde{a}$, this gives
\begin{equation}\label{eq:s_bound}
    s\approx 2 + \frac{4}{\pi} \frac{\mathcal{C}_z +\mathcal{C}_x(2 - \sin^2 \theta_L)}{\mathcal{C}_m}\approx 2.76-0.1 \sin^2 \theta_L\,.
\end{equation}
The last equality is independent of the local environment density and is approximately independent of orbital radius for $r\gtrsim 50 R_M$; remarkably, it holds for any $\gamma_{\rm sp} \in \left( 1.5, 3\right)$ to an error $\lesssim 1.5 \%$.
As we discuss in more detail in \cref{sec:discussion,sec:conclusions}, this spin-evolution parameter differs considerably from the typical values found in other astrophysical environments (which often exhibit $s\leq2$); thus, it could, in principle, be explored as a distinctive signature of dense (collisionless) DM environments.

\Cref{fig:alpha_mdot} shows the co-evolution of a BH's spin magnitude with its mass over astrophysical timescales, for a quasi-circular IMRI within a DM spike environment with fiducial parameters \eqref{eq:fiducial}. The numerical solution is obtained from integrating \cref{eq:s}, using \cref{eq:approx_results,eq:dJdt_approx}. {This is then compared to the corresponding solution of \cref{eq:s_bound} at leading order in $\tilde{a}$, which closely matches the numerical solution. We also compare the solution to the case $s=2$, which corresponds to spin-down from mass growth at constant angular momentum.

As before, we show the spin magnitude evolution as a function of the associated time to merge, assuming a quasi-circular inspiral driven by GW emission and accretion drag (implemented as in~\cite{karydas2024sharpeningdarkmattersignature}), with the fiducial parameters in \cref{eq:fiducial}. We also show the relative change in spin and mass over a timescale of a few years before the merger. 

In optimistic scenarios, LISA could measure the companion's mass of an IMRI to a precision $\lesssim10^ {-3}$~\cite{Huerta:2011zi}, comparable to the changes shown in \cref{fig:alpha_mdot}. %Contrarily, 
The secondary's spin in IMRIs is instead not well constrained by LISA observations~\cite{Huerta:2011zi}. The observation of our fiducial system for the last 4~yrs before merger in LISA would lead to a signal-to-noise ratio $\sim 500$ for a luminosity distance $\lesssim 1 {\rm\, Gpc}$~\cite[Fig. 5]{Coogan_2022}.\\

\paragraph*{\textbf{The effect of eccentricity.}}

While we have ignored the effect of orbital eccentricity in our analysis, it can, in principle, influence our results in two ways. First, by enhancing GW emission, eccentric orbits allow the companion to inspiral from larger separations for a fixed timescale \cite{Maggiore:2007ulw}. Second, the accretion and dynamical friction rates are modulated as the companion traverses regions of varying density and velocity along its orbit. 

We proceed then to calculate the spin evolution parameter, taking into account generic orbits with orbital eccentricity $e$. For simplicity, and given the subdominant contribution of tilt misalignment (cf. \cref{eq:s_bound}), we focus on the case where $\theta_L = 0$, and we perform the time-averaging of $\dot{m}$ and $\dot{J}_z$ using,
\begin{equation}
    \left< \cdots \right> = \left(1-e^2\right)^{3/2} \int_0^{\pi} \cdots \frac{1}{\pi} \left( 1 + e \cos \nu \right)^{-2}\, \mathrm{d}\nu\,,
\end{equation}
with the separation $r(\nu)$ and velocity $u(\nu)$ are calculated as in \cite{karydas2024sharpeningdarkmattersignature}. 

We find the spin-evolution parameter $s(e)$ to be approximately constant, with a very modest increase between 0.03\% and 2\%, as we vary the spike's slope $\gamma_{\mathrm{sp}} \in (1.5, 3)$, and for eccentricity up to $e = 0.8$.\\

\paragraph*{\textbf{The effect of feedback.}}
The feedback induced on the DM distribution is non-negligible for an IMRI; so the distribution function of the DM particles should be evolved alongside the companion's orbit \cite{Kavanagh_2020,Coogan_2022}. In our analysis, we account for the suppression of dynamical friction due to feedback \cite{Cole_2023,Kavanagh_2020}, but neglect its effect on the background density profile. We check its importance by incorporating \cref{eq:dadt}, with \cref{eq:approx_results,eq:dJdt_approx}, in the formalism of \cite{karydas2024sharpeningdarkmattersignature}, using \texttt{HaloFeedback} \cite{HaloFeedback} to evolve the distribution function. We find that feedback does not impact the spin-evolution parameter $s$ for the quasi-circular case and induces a mild reduction at larger eccentricities; here, we report $\lesssim 3\%$ for $e = 0.4$ with the maximum deviation during the transient depletion phase (cf. \cite{Kavanagh_2020}).

\subsection{Spin measurements from GWs}

The changes in spin vector imparted by DM accretion over a few years are exceedingly small, even in environments as dense as DM spikes. However, these effects are stronger over astrophysical timescales $\gtrsim 10^6\, {\rm yrs}$ (c.f., \cref{fig:alpha_mdot}), and potentially strong enough to be probed through (late inspiral) spin measurements from GW observations.

In post-Newtonian (PN) theory, the leading effect from BH spins arises from a spin-orbit coupling at $1.5$PN order \cite{Tanaka_1996,Maggiore:2007ulw,Blanchet:2013haa}. The spin components aligned with the orbital angular momentum enter the energy-flux balance and phase evolution, whereas in-plane (orthogonal) components drive precession, causing a time-dependent modulation of the waveform. 
In asymmetric binaries like E/IMRIs, the primary's spin is typically the only one well constrained by GW observations, as the mass-ratio suppresses the effects from the secondary's spin. 
Among the secondary's spin components, the projection along the orbital angular momentum, $\chi\equiv \tilde{a} \cos \theta_L$, is the most likely to be measurable, since it produces a secular phase shift in the waveform.
Indeed, Ref.~\cite{Huerta:2011zi} showed that for IMRIs with mass-ratios $q\gtrsim 10^{-3}$, LISA could measure the secondary's aligned spin to a precision $\lesssim 0.1$ if both binary components are rapidly rotating. On the other hand, for EMRIs it is usually thought that measuring the spin of the companion will be difficult \cite{Piovano_2021,Piovano_2020,Skoup__2022,burke2024accuracyrequirementsassessingimportance,cui2025secondaryspinsextrememass,cui2025secondaryspinsextrememass}.

As demonstrated in the previous sections, DM accretion onto an IMRI's secondary induces both spin-alignment towards the orbital plane and spin-down, with the latter being dominant; together, they act to reduce the magnitude of $\chi$. In \cref{fig:bounds}, we combine these results to show the secondary's aligned spin component at merger for different evolution timescales and DM spike normalizations, assuming an initially maximal spin $\chi=1$. In other words, the secondary's aligned spin of an IMRI evolving over a given timescale within a DM spike of specified density \emph{must} lie below the corresponding colored region. We also illustrate the impact of the DM spike slope, which becomes significant only for short evolution timescales.

\begin{figure}[hb!]
\centering
\hspace*{-0.05\columnwidth}
    \includegraphics[width=0.9\columnwidth]{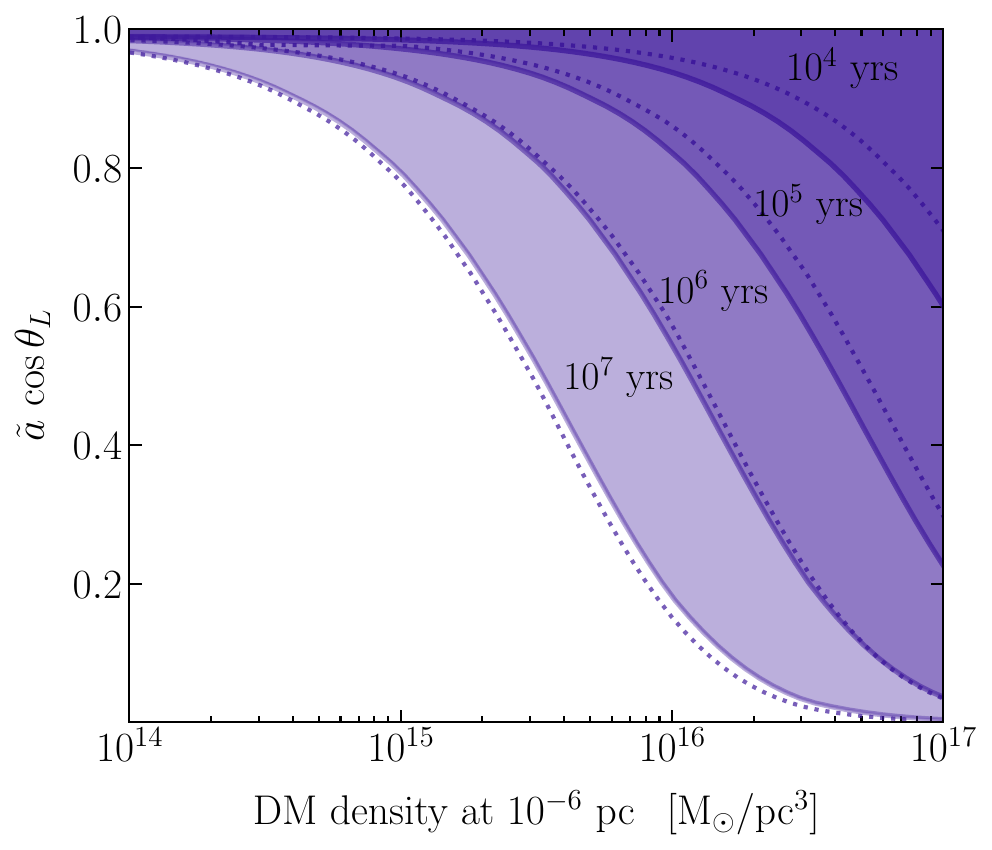}
    \caption{\textbf{Forbidden region for secondary's aligned-spin component at merger, $\bm{\chi=\tilde{a}}$\bf{cos}$\bm{\theta_L}$.
    }
    The secondary's spin of an IMRI evolving over a given timescale within a DM spike of specified density must lie below the corresponding colored region. The results correspond to the fiducial slope $\gamma_{\rm sp}=7/3$; dashed lines show results for $\gamma_{\rm sp}=2$.
    \label{fig:bounds}}
\end{figure}

We note that, for rapidly spinning primaries, spin measurements from GW observations with LISA could be sufficiently precise (e.g., \cite{Huerta:2011zi}) to exclude the presence of dense DM environments ($\gtrsim 10^{16}\, \mathrm{M_\odot/pc^3}$), if a rapidly spinning secondary is observed---assuming such IMRI evolves over timescales of at least $\gtrsim 10^6\,{\rm yrs}$. 

Even in dense stellar clusters, hierarchical mergers are not expected after the IMRI forms; for $r\lesssim 10^{-3}\,\mathrm{pc}$, the timescale for tertiary capture is larger than a Hubble time \cite{Quinlan:1996vp}. Accretion from intracluster gas could lead to $\dot{m}/m\lesssim 10^{-10}(m/10\mathrm{M_\odot})\,\mathrm{yr^{-1}}$ for an IMRI at the centre ($r\lesssim 10^{-3}\,\mathrm{pc}$) of a gas-rich cluster~\cite{Bobrick:2025rqh}; this results in a negligible spin-up over $10^7\,\mathrm{yrs}$~\cite{Bardeen:1970zz}.\footnote{For reference, the Eddington rate $\dot{m}_{\rm Edd}/m\sim 10^{-8} \, \mathrm{yr^{-1}}$ results in a spin up of $\Delta \tilde{a}\lesssim 0.3$ over $10^7\,\mathrm{yrs}$~\cite{Bardeen:1970zz}.} Our method is robust, as we do not expect any spin-up from astrophysical processes over the timescale of the inspiral---the observation of rapidly spinning IMRI companions provides a means to probe the existence of dense DM environments without relying on GW dephasing~\cite{Cole:2022yzw, Vicente:2025gsg}. 

\section{Discussion} \label{sec:discussion}
In this section, we discuss several caveats associated with our calculations, focusing in particular on the assumptions made in modeling the DM distribution and on the applicability of our results to systems beyond the fiducial case. We then address the robustness of our predictions for the mass–spin co-evolution in the presence of potential degeneracies.\\

\paragraph*{\textbf{Spike survival at large radii.}}
For CDM, it is in principle possible that DM overdensities get disrupted by mergers with massive BHs. Disruption is highly sensitive to the mass ratio: only equal-mass mergers substantially destroy the spike, while lower-mass companions have a comparatively mild effect. Thus, survival is expected provided the central BH did not undergo an equal-mass merger after obtaining its present mass~\cite{Merritt_2002}. 

Spikes can also be substantially depleted by the gravitational heating induced by a stellar distribution \cite{Bertone_2005,Merritt_2007,Shapiro_2022}. We note however that such depletion is only effective in regions co-occupied by a stellar population \cite[cf. `orbit-averaging' of the Fokker-Planck equation]{2013degn.book.....M} and is insensitive to the specific stellar masses as long as a mass hierarchy is satisfied, $m_{\rm DM}/m_* \ll 1$ \cite{PhysRevLett.92.201304}.

The extent of stellar-induced heating around BHs remains uncertain. 
For the only well-characterized system, the Galactic Center, the heating of the DM spike is expected to be effective only down to radii of $\sim6\times10^{-4}\,\mathrm{pc}$, corresponding to the periastron of the S2 star~\cite{S2_star}, which roughly bounds the size of the S-cluster~\cite{Ali_2020}. 
At smaller radii, only a sparse population of $\mathcal{O}(10)$ unresolved stars is expected~\cite{gravityconstraints}, insufficient to generate significant heating through dynamical-friction feedback~\cite{Kavanagh_2020}. 
These considerations suggest that the density spike remains essentially unperturbed at the smaller separation scales ($\sim10^{-6}\,\mathrm{pc}$) relevant for this study. 
We will return to the implications of both disruption channels in future work.\\

\paragraph*{\textbf{Spherical symmetry and isotropy.}}

To keep the problem trackable, we approximated the DM velocity distribution as spherically symmetric and isotropic. In realistic configurations, these assumptions are violated at the center: within $\sim 10 \, R_M$, a Kerr primary focuses DM equatorially \cite{Ferrer_2017} and loss-cone depletion generates tangential anisotropy \cite{Sadeghian_2013}. Such effects, however, are negligible at the large separations relevant to this work.

DM spikes also form with some anisotropy \cite{Gondolo_1999,bertone2024darkmattermoundsrealistic}. To assess its impact on our predictions, we compute the spin parameter $s$ using \cref{eq:s_bound} with coefficients from \cref{eq:cj0,eq:cm}, calculated using the anisotropic distribution of \citet{Gondolo_1999}. Utilizing Monte Carlo integration over velocity space, we find $s \simeq 2.72$ for $\theta_L = 0$, which is only $\sim1\%$ different compared to the isotropic case.\\

\paragraph*{\textbf{Systems of interest.}}

While our focus was on DM spikes around asymmetric BH binaries of the E/IMRI type, the framework is general and can be applied to binaries within other collisionless environments as long as they are described by a distribution function. Additionally, it can be applied to isolated BHs moving through a collisionless environment. The regime of interest in this work lies firmly within the weak-field limit, where the ignored relativistic corrections are negligible. Our results can be extended, however, to the relativistic regime by implementing a relativistic distribution function (e.g. \cite{Sadeghian_2013}) in \cref{eq:accretion_in_kerr,eq:angtum_change}, and by evolving the geodesic constants of motion of the companion as~in~\cite{Vicente:2025gsg}, we note however that the co-evolution happens at the weak field.

To identify systems beyond the fiducial where the studied effects are astrophysically relevant, we adopt the practical criterion $\Delta m/m \gtrsim 0.1$, which corresponds to an $\sim 30\%$ change in the spin for typical parameter choices. Noting that most mass is accumulated early in the inspiral, we approximate $\Delta m \simeq \dot{m}(r_0) \, t_{\to\mathrm{merge}}$, where $r_0$ results in a merger after $t_{\to\mathrm{merge}}$. Restricting to the regime $q \gtrsim 10^{-5}$, where the inspiral is dominated by gravitational-wave emission and the spike-induced orbital feedback is subdominant, we find the scaling
\begin{equation}
    \frac{\Delta m}{m} \approx \frac{1}{5} \frac{\rho_6~10^{-16}}{\mathrm{M}_\odot/\mathrm{pc}^3} \!\left(\frac{M}{10^4\,\mathrm{M}_\odot}\right)^{\frac{7-6\gamma_\mathrm{sp}}{8}} \!\!\! \left(\frac{q t_{\to \rm merge}}{1000~\mathrm{yr}}\right)^{\frac{9-2\gamma_\mathrm{sp}}{8}}\!\!\!\!\!\!\!\!\!, \nonumber
\end{equation}
with $\rho_6\equiv \rho(10^{-6}\,\mathrm{pc})$.

This expression makes two points explicit: (i) for fixed mass ratio and spike normalization, increasing the primary mass suppresses accretion (owing to the lower DM density at the corresponding larger scales), and (ii) $\Delta m/m$ decreases with decreasing mass ratio. The latter implies that our results are most relevant in the IMRI regime. Further extending this criterion to smaller $q$ or beyond the quasi-circular approximation requires a full numerical treatment that also includes spike-induced energy losses.\\

\paragraph*{\textbf{Robustness against degeneracies.}}

To assess the robustness of our results, we examine potential degeneracies with other spin-altering mechanisms.

The closest conceptual analog is the treatment by \citet{Hughes_2003}, where the spin of an isolated intermediate-mass (or massive) BH evolves through a sequence of minor mergers: smaller compact objects are dynamically captured, inspiral via GW emission, and deposit their remaining energy and angular momentum at the last stable orbit. The cumulative effect of isotropically oriented captures leads to spin-down with $s \approx 2.4$ \cite{Hughes_2003}, close to the value found here ($s \approx 2.8$), and negligible coherent precession. However, during a GW-driven IMRI, the orbital velocity of the intermediate-mass BH companion is sufficiently large that the probability of capturing additional compact objects over the inspiral timescale is vanishingly small, even in dense stellar environments. Consequently, the spin evolution in that scenario is governed exclusively by continuous DM accretion rather than by discrete merger events.

Gas accretion offers another spin-altering channel~\cite{Font_1999,ricarte2023recipesjetfeedbackspin,2005ApJ...620...59S,Bogdanovi__2007,Secunda_2020,Cenci_2020,bartos2025accretionneedblackhole,Gerosa_2015,Li__2022,Tagawa_2020}. Embedded BHs in accretion disks are typically spun up efficiently and aligned with the disk angular momentum \cite{Bogdanovi__2007,bartos2025accretionneedblackhole}, in contrast to the DM-driven case, where they are spun down and aligned with the orbital plane. This is also distinctly different from the relativistic Lense-Thirring and de Sitter effects, which induce a gyroscopic precession instead \cite{Gangardt_2021}.

\section{Conclusions}
\label{sec:conclusions}

In this work, we analyzed how BHs accrete from collisionless DM environments, like DM spikes~\cite{Gondolo_1999, Sadeghian_2013, Ferrer_2017}. We focus on the evolution of their mass and spin vector (both the magnitude and direction). Larger spin parameters $\tilde{a}$ lead to smaller accretion rates (\cref{fig:accretion_mass_force}), but stronger torques (\cref{fig:torque_alpha}) that spin-down and secularly align the companion's spin to the orbital plane. The main implications of our work are:
\begin{enumerate}
    \item The accretion process imprints a characteristic correlation in mass-spin of BHs, with spin-evolution parameter $s\approx 2.8$ (cf. \cref{eq:s}),
    independent of local density
    and spike's slope $\gamma_{\rm sp}$ (or, equivalently, from velocity distribution). The universality of the result may allow us to combine mass-spin measurements of IMRI secondaries from future GW observations to unveil the presence of dense DM environments; we leave a detailed exploration for future work. The resulting spin-evolution parameter is thus much larger than  
    in the case of (collisional) astrophysical environments, which typically lead to $s\leq 2$--for example, thin~\cite{Bardeen:1970zz, Thorne:1974ve} or slim~\cite{Sadowski:2011ka} disks, Bondi-Hoyle-Lyttleton accretion~\cite{Shapiro:1976, Ruffert:1995}, or hierarchical mergers~\cite{Fishbach:2017dwv, Gerosa:2021mno}.
    \item Collisionless DM accretion causes IMRI secondaries to spin down on relevant astrophysical timescales ($\gtrsim 10^6\,\mathrm{yrs}$, for the fiducial parameters in~\cref{eq:fiducial}). Other astrophysical processes (e.g., gas accretion or hierarchical mergers) are not expected to spin up the IMRI secondary during the inspiral. Thus, the observation of rapidly spinning IMRI companions can, in principle, rule out the presence of dense DM environments (cf. \cref{fig:bounds}), providing complementary constraints to those arising from dynamical friction.
\end{enumerate}

\section*{Acknowledgments}
We gratefully acknowledge the support of the Dutch Research Council (NWO) through an Open Competition Domain Science-M grant, project number OCENW.M.21.375.

\onecolumngrid
\appendix

\section{Particle capture in Kerr} \label{app:bcr_calculate}
In this work, we operate within the Kerr metric in Boyer–Lindquist coordinates to describe the spinning BH companion,
\begin{equation}
    ds^2 = -\left( 1 - \frac{r_sr}{\Sigma} \right) c^2 dt^2 + \frac{\Sigma}{\Delta}dr^2 +\Sigma \,d\theta^2 +\left( r^2 +a^2 +\frac{r_s r a^2}{\Sigma} \sin^2\theta \right)\sin^2\theta\,d\phi^2 -\frac{2r_s r a}{\Sigma} \sin^2\theta \, c dt d\phi\,,
\end{equation}
where $a = J/m/c$ is a spin-related length-scale such that $\tilde{{a}}~=~a/R_m \in [0, 1)$, $R_m = Gm/c^2$, and $\Sigma = r^2 +a^2\cos^2\theta$, $\Delta = r^2 -r_s r +a^2$. This line element permits the Lagrangian $\mathcal{L} = -g_{\mu\nu} \dot{x}^\mu \dot{x}^\nu$ where $g_{\mu\nu}$ are the metric's contributions to the differentials $dx^\mu dx^\nu$. Subsequently, the orbits of massive particles in the rest frame of the BH are described by a set of four coupled differential equations ($t, r, \theta, \phi$), and three integrals of motion, namely specific energy $\calE$, angular momentum on the axis of rotation $h_z$, and Carter's constant $\mathcal{K}$ \cite{PhysRev.174.1559}. For radial motion we have \cite{Misner:1973prb,Kapec_2019},
\begin{equation}
    \dot{r}^2 = \frac{R(r)}{\Sigma^2}\,, \qquad \text{with} \qquad R(r) := \left[ ah_z -\frac{\calE}{c}\left(r^2 +a^2 \right)\right]^2 - \Delta \left[ \mathcal{K} +c^2 r^2 +\left( h_z -\frac{a \calE}{c} \right)^2 \right] \,.
\end{equation}

\paragraph{\textbf{Constants of motion.}}
We calculate the constant's of motion at infinity by relating them to a particle approaching with an impact parameter $\bm{b}$ and velocity $\bm{u}$. Specifically for the specific energy $\calE = \gamma c^2 $ and angular momentum $h = \gamma b u$ where $\gamma = 1/\sqrt{1-\beta^2}$ is the Lorentz factor and $\beta = u/c$. Moreover the orbit approaches on a planed inclined by an angle $i$ and with an incidence angle (with the rotation axis $\theta_V$). As it has been pointed out before \citet{Glampedakis_2002}, the inclination angle in Kerr geometry may be identified in various ways \cite{Gair_2006,PhysRevD.108.084065,Shibata_1995,Ryan_1995,Komorowski_2010,Komorowski_2012,Grossman_2012,Hod_2013} with different properties and relating to the context and problem at hand. In our analysis, we adopt a definition that ensures consistency with the flat-space derivation at spatial infinity; we require that $\cos i = h_z/h = \sin\theta_V \cos\chi$. Here the total angular momentum $h$ is naturally related to the rest of the integrals of motion and the incidence polar angle at infinity $\theta_\infty = \pi -\theta_V$ through $\mathcal{K} = h^2 -h_z^2 -a^2 (\calE^2/c^2 -c^2)\cos^2\theta_V$. While the second term introduces a breaking of spherical symmetry \cite{ROSQUIST_2009}, the effect is negligible in many scenarios, as $h^2-h^2_z \propto b^2 \gg a^2$, except during highly energetic encounters. In this work we focus specifically on encounters without very relativistic velocities.\\

\paragraph{\textbf{Finding the critical impact parameter.}}
At the critical threshold of particle capture, the inversion point of the particle's radial trajectory corresponds to an extrema of the radial potential $R(r)/\Sigma^2$. Since the potential $\Sigma$ has no singularities except equatorially at $r = 0$, the extremum condition for the radial motion requires that both the function $R(r)$ and its first derivative $R'(r)$ with respect to $r$ vanish simultaneously. Since $R(r)$ and by extension $R'(r)$ are polynomials in $r$, the condition for their shared roots is satisfied whenever their resultant vanishes. This is equivalent to the vanishing of the discriminant $\Delta(b_\mathrm{cr}, \tilde{a}, \cos i, \cos\theta_V, \beta)$ of the $R(r)$ polynomial. The critical impact parameter can then be rapidly inverted numerically for general orbits around the BH. However, a downside to this method is the possibility of non-physical solutions emerging that correspond to a point of closest approach $R(r_0) = 0$ inside the event horizon. This must be checked by the root-finding algorithm. The polynomial whose discriminant is calculated is,
\begin{equation} \label{eq:bcr_polynomial}
    1 + \frac{2}{\beta^{2} \gamma^{2}} t
    + \left(\tilde a^{2} (1 +\cos^2\theta_V) - \tilde b^{2} \right) t^2
    + 2\left(\tilde b^{2} - 2 \cos i \frac{\tilde a}{\beta} \tilde b + \frac{\tilde a^{2}}{\beta^2} -a^2 \cos^2\theta_V\right) t^3
    + \left( \tilde a^2 \cos^2\theta_V - \tilde b^2 \sin^2 i \right) \tilde a^2 t^4 = 0,
\end{equation}
where $t = 1/\tilde{r}$ and $\tilde{\cdots} = \cdots/R_m$. For non-spinning BHs, one simply recovers the well-known formula for the critical impact parameter \cite{PhysRevD.14.3251}
\begin{equation}
  b_\mathrm{cr}(\beta) = \frac{R_m}{\sqrt{2} \beta^2} \frac{\left( \sqrt{8\beta^2+1}+4\beta^2 -1 \right)^{3/2}}{\sqrt{8\beta^2+1} -1}\,,
\end{equation}
and may additionally recover the case of photon capture in the limit $\beta \approx 1$ in equatorial \cite{Iyer_2009} and polar incidence \cite{Dolan_2010}.

\section{Regarding transfer of linear momentum}
Unsurprisingly, repeating the numerical experiment in \cref{ssec:mass_force} for the force $\bm{F}_\mathrm{acc}$ that arises from the accretion of linear momentum associated with the in-falling particles by altering \cref{eq:accretion_in_kerr} as in \cite{karydas2024sharpeningdarkmattersignature}, we find that the drag component is fitted by an equation similar to \cref{eq:approx_results},
\begin{equation}
    F_\mathrm{AF} = -\bm{F}_\mathrm{acc} \cdot \bm{u} \approx \rho \, \sigma(\tilde{a}, u) \,u^2 \, \mathcal{C}_\mathrm{acc}\,,
\end{equation}
where $\mathcal{C}_\mathrm{acc}$ is equivalent to $\mathcal{C}_m$ and discussed in more detail in \cite{karydas2024sharpeningdarkmattersignature}. While other components vanish for the Schwarzschild companion \cite{karydas2024sharpeningdarkmattersignature}, it is not generally true for arbitrary spins and non-wind-like distributions as in \citet[Eq.14]{Dyson:2024qrq}. A new component emerges perpendicular to the velocity and lying in the plane between that and the spin axis, e.g. an accretional Magnus force \cite{1961JFM....11..447R}, which may tilt the orbital plane or induce precession depending on the relative orientation of these vectors. However, we find it to be very small compared to $F_\mathrm{AF}$ and indeed using the leading order in $\tilde{a}$ of \cref{eq:fitb2}, we can show that its leading term is proportional to $\tilde{a}^2 \beta^2$.

\section{Spin expansions of cross-sections} \label{app:expansions}
To evaluate the integrals of $b^2_\mathrm{cr}$ and $b^3_\mathrm{cr} \sin \chi$ in \cref{eq:accretion_in_kerr} and \cref{eq:spin_djdt} respectively, we employ the discriminant $\Delta(\beta, \tilde{b}, \tilde{{a}}, A)$ associated with the quartic polynomial given in \cref{eq:bcr_polynomial}. By expanding $b^2_\mathrm{cr}$ and $b^3_\mathrm{cr}$ as Taylor series around $\tilde{a} = 0 $ we obtain two algebraic expressions that are valid up to the reported powers of the spin parameter below. The coefficients of the expansions are functions of the ``inclination'' coefficient $A(\chi)$ and the relative velocity $\beta = V/c$ (where $\theta_\mathrm{V} \to \theta$, $b_\mathrm{cr}(\tilde{a}=0) \to b_0$  for brevity). By performing the integration over a complete circle in the plane of the encounter, the resulting expressions
\begin{align}
    \frac{1}{2 b^2_0} &\int_0^{2\pi} b^2(\chi) \, \mathrm{d}\chi = \pi - \tilde{a}^2 F_2(\beta, \sin\theta) - \tilde{a}^4 F_4(\beta, \sin\theta) +\mathcal{O}(\tilde{a}^6)\,, \label{eq:fitb2}
    \\
    \frac{-1}{3 b_0^3} &\int_0^{2\pi} b^3(\chi) \cos(\chi) \, \mathrm{d}\chi = \tilde{a} F_1(\beta, \sin\theta) +\tilde{a}^3 F_3(\beta, \sin\theta) +\mathcal{O}(\tilde{a}^5)\,,\label{eq:fitb3}
\end{align}
are considerably simplified due to symmetry which eliminates at minimum the first three odd and even contributions. The remaining terms' angular ($\sin\theta$) and velocity dependence is separable in the form
\begin{align}
    F_1(\beta, \sin\theta) &= l_1(\beta) \sin\theta \\
    F_2(\beta, \sin\theta) &= k_2(\beta) + l_2(\beta) \sin^2\theta\,, \\
    F_3(\beta, \sin\theta) &= l_3(\beta) \sin\theta + m_3(\beta) \sin^3\theta\,, \\
    F_4(\beta, \sin\theta) &= k_4(\beta) + l_4(\beta) \sin^2\theta + m_4(\beta) \sin^4\theta\,,
\end{align}
where $k,l,m$ are smooth, monotonic but algebraically complicated functions of $\beta$. Despite their complexity, we find that each can be closely approximated with high accuracy (with goodness of fit $1-R^2 \leq 10^{-5}$) by a polynomial expression:
\begin{equation}
    \begin{aligned}
        k_2 &\approx \frac{\pi}{16} \left( 1 +\frac{859}{379} \beta^2 -\frac{839}{373}\beta^3 +\frac{470}{617} \beta^4\right)\,,
        \\
        k_4 &\approx \frac{\pi}{128} \left( 1+\frac{106}{47} \beta^2 -\frac{86}{59} \beta^3 +\frac{4}{13}\beta^4 \right)\,,
        \\
        m_3 &\approx -\frac{\pi}{128} \left(1 -\frac{16}{301} \beta -\frac{194}{85} \beta^2 +\frac{721}{202} \beta^3 - \frac{230}{63}\beta^4 +\frac{186}{131}\beta^5\right)\,,
        \\
        m_4 &\approx -\frac{9\pi}{2048} 
        \left( 1 +\frac{1}{95} \beta -\frac{31}{21} \beta^2 +\frac{769}{279} \beta^3 -\frac{475}{169} \beta^4 +\frac{729}{722} \beta^5 \right) \,,
    \end{aligned} \quad \,
    \begin{aligned}
        l_1 &\approx \frac{\pi}{4} \left(1 +\frac{155}{91} \beta^2 -\frac{178}{99} \beta^3 +\frac{7}{11} \beta^4\right)\,,
        \\
        l_2 &\approx \pi \left(0 - \frac{209}{956} \beta^2 +\frac{242}{797}\beta^3 -\frac{157}{978}\beta^4 +\frac{10}{501} \beta^6 \right)\,,
        \\
        l_3 &\approx \frac{\pi}{64} \left( 1 -\frac{50}{19}\beta^2 +\frac{64}{25} \beta^3 -\frac{441}{475} \beta^4 \right)\,,
        \\
        l_4 &\approx \frac{3\pi}{256} \left(1 -\frac{569}{215}\beta^2 +\frac{936}{371} \beta^3 -\frac{65}{74} \beta^4\right)\,.
    \end{aligned} \nonumber
\end{equation}

By substituting \cref{eq:fitb2,eq:fitb3} into \cref{eq:approx_results,eq:dJdt_approx} respectively, we find the orbit-averaged $\dot{m}$ and $\dot{J}_z$ for a quasi-circular inspiral as a function of the companion's spin parameter $\tilde{a}$ and the angle between the spin axis and orbital angular momentum axis,
\begin{align} \label{eq:spin_evolution_eqs}
    &\frac{\dot{m}}{\rho \, u\, b_\mathrm{cr}^2 \, \mathcal{C}_m} = \pi -\tilde{a}^2 \left(k_2 +l_2\right)-\tilde{a}^4 \left(k_4 +l_4 +m_4\right)
        + \frac{\tilde{a}^2 \sin^2 \theta_L}{2} \left( l_2 +l_4 +2\tilde{a}^2 m_4 \right)
        - \frac{3}{8}\tilde{a}^4 m_4 \sin^4 \theta_L\,,\\
    &\frac{\dot{J}_z}{\rho \, u^2\, b_\mathrm{cr}^3} = -\left(\mathcal{C}_z +2\mathcal{C}_x\right) \left(1 + \tilde{a}^2 \frac{l_3 +m_3}{l_1} \right) 
    + \sin^2 \theta_L \left[ \mathcal{C}_x + \tilde{a}^2 \frac{2\mathcal{C}_x l_3 +m_3 \left( \mathcal{C}_z +4\mathcal{C}_x\right)}{2l_1} \right] 
    - \frac{3}{4}\tilde{a}^2 \frac{m_3}{l_1} \mathcal{C}_x \sin^4 \theta_L\,. \label{eq:spin_evolution_eqs2}
\end{align}

\twocolumngrid
\bibliography{bibliography}
\end{document}